\documentclass[12pt,preprint]{aastex}

\usepackage{emulateapj5}
\usepackage{epsf}
\usepackage{graphics}

\newcommand{\kms}{km\,s$^{-1}$}
\newenvironment{inlinefigure}{
\def\@captype{figure}
\noindent\begin{minipage}{0.999\linewidth}\begin{center}}
{\end{center}\end{minipage}\smallskip}

\slugcomment{Accepted to the Astrophysical Journal}

\shorttitle{Disk Galaxy Stellar and Dark Masses}
\shortauthors{Conselice et al.}

\def\deg{$^{\circ}\,$}
\def\solm{M$_{\odot}\,$}
\def\kms{km s$^{-1}$}

\begin{document}

\title{Evolution of the Near-Infrared Tully-Fisher Relation: Constraints on the Relationship Between the Stellar and Total Masses of \\ Disk Galaxies since $z \sim 1$}

\author{Christopher J. Conselice$^{1,2}$, Kevin Bundy$^{1}$, Richard S. Ellis$^{1}$, Jarle Brichmann$^{3,4}$, Nicole P. Vogt$^{5}$, Andrew C. Phillips$^{6}$}
\altaffiltext{1}{California Institute of Technology, Mail Code 105-24, Pasadena
CA 91125}
\altaffiltext{2}{NSF Astronomy \& Astrophysics Postdoctoral Fellow}
\altaffiltext{3}{Max-Planck-Institut f{\"u}r Astrophysik,
Karl-Schwarzschild-Str. 1, 85740 Garching bei M{\"u}nchen}
\altaffiltext{4}{Centro de Astrof{\'\i}sica da Universidade do Porto, Rua das
Estrelas - 4150-762 Porto, Portugal}
\altaffiltext{5}{New Mexico State University, Astronomy Department, Las Cruces, NM 88003}
\altaffiltext{6}{Lick Observatory, University of California, Santa Cruz, CA 95064}

\begin{abstract}

Using a combination of Keck spectroscopy and near-infrared
imaging, we investigate the K-band and stellar mass Tully-Fisher relation 
for 101 disk galaxies at 0.2 $<z<$1.2, with the goal of placing the first 
observational constraints
on the assembly history of halo and stellar mass.
Our main result is a lack of evolution in either the
K-band or stellar mass Tully-Fisher relation from $z = 0 - 1.2$.
Furthermore, although our sample is not statistically complete, we
consider it suitable for an initial investigation of how the fraction of total
mass that has condensed into stars is distributed with both
redshift and total halo mass.  We calculate stellar masses from 
optical and near-infrared photometry
and total masses from maximum rotational velocities and disk scale
lengths, utilizing a range of model relationships derived analytically and
from simulations.   We find 
that the stellar/total mass distribution and stellar-mass Tully-Fisher
relation for $z>$0.7 disks is
similar to that at lower redshift, suggesting that baryonic mass is accreted
by disks along with dark matter at $z < 1$, and
that disk galaxy formation at $z < 1$ is hierarchical in nature.
We briefly discuss the evolutionary trends expected in conventional structure
formation models and the implications of extending such a study to
much larger samples.

\end{abstract}

\section{Introduction}

In the currently popular hierarchical picture of structure
formation, galaxies are thought to be embedded in massive dark
halos.  These halos grow from density fluctuations in the early
universe and initially contain baryons in a hot gaseous phase.
This gas subsequently cools and some fraction eventually condenses
into stars. Much progress has been made in observationally
delineating the global star formation history and the resulting
build-up of stellar mass (e.g., Madau et al. 1998, Brinchmann \& Ellis
2000; Dickinson et
al. 2003; Bundy, Ellis \& Conselice 2005). 
However, many of the physical details, particularly the
roles played by feedback and cooling, essential for a full
understanding of how galaxies form, remain uncertain. 
Models (e.g., van den Bosch 2002; 
Abadi et al. 2003) have great predictive power in
this area but only by assuming presently-untested prescriptions
for these effects. Obtaining further insight into how such
processes operate is thus an important next step not only in
understanding galaxy evolution, but also in verifying the utility
of popular models, as well as the hierarchical concept itself.   One approach
towards understanding this issue is to trace how the stellar mass in
galaxies forms in tandem, or otherwise, with its dark mass.

The first step in this direction began with studies of scaling
relations between the measurable properties of disk galaxies,
specifically the relation between luminosity and maximum rotational
velocity (Tully \& Fisher 1977).  Studies utilizing roughly a thousand
spiral galaxies have revealed a tight correlation between absolute
magnitude and the maximum rotational velocity for nearby galaxies
(Haynes et al. 1999).  The limited data at high redshift
suggests the TF relation evolves only modestly, equivalent to at
most 0.4 - 1 magnitudes of luminosity evolution to $z \sim 1$
(Vogt et al. 1997; Ziegler et al. 2002; Bohm et al. 2004).  How the
Tully-Fisher relation evolves with redshift is still controversial,
although it appears that fainter disks evolve the most (Bohm et al. 2004), 
and that selection effects are likely dominating 
the differences found between various studies.  Furthermore, it has
been difficult for modelers to reproduce the Tully-Fisher relation
to within 30\% (e.g., Cole et al. 2000), making it an important constraint 
on our understanding of the physics behind galaxy formation.

Unfortunately, any interpretation of the TF relation is
complicated by the fact that both luminosity and virial mass might be
evolving together. A more physically-motivated comparison would be 
between stellar and virial mass. Not only does this relation break
potential degeneracies in the TF technique, but it also samples
more fundamental quantities.
In this paper we begin this task by investigating the evolution in the
fraction of the total
mass that is in a stellar form. This can be accomplished with some
uncertainty by contrasting the {\it stellar mass} of a galaxy with
its {\it halo mass}. We selected disk galaxies for our initial
study since these two quantities can be effectively probed
observationally for these galaxies with various assumptions 
(e.g., van den Bosch 2002; Baugh et al. 2005).

This paper presents the first study of the near-IR TF relation, as well
as a comparison between stellar and halo masses, for 101 disk galaxies
within the redshift range $0.2 < z < 1.2$ drawn mostly from
the DEEP1 redshift survey (Vogt et al. 2005). Our goal is to address several 
questions relating to the mass assembly history of disks. As our sample is not
formally complete in any sense, we cannot derive general
conclusions concerning the history of {\em all} present-day disks.
However, we can determine whether the disks selected from the DEEP1
survey in the sampled redshift range are still accreting matter
and converting baryons into stellar disks at a significant rate.
We construct stellar mass Tully-Fisher and stellar mass/halo mass
relation for our sample and find that there is little evolution
in either from $z \sim 0 - 1.2$.  This suggests that the dark and stellar
components of disk galaxies grow together during this time.

The paper is organized as follows: \S 2 contains a description of
the sample including the fields used and the different data
products and a discussion on quality control.  \S 3
describes how various quantities, such as the halo and stellar
masses, are derived from the basic data.  \S 4 presents our
results and \S 5 summarizes our conclusions.   We
assume the following cosmology throughout this paper: H$_0$ = 70 km s$^{-1}$
Mpc$^{-1}$, $\Omega_{\Lambda}$ = 0.7 and $\Omega_{\rm m}$ = 0.3.

\section{Data}

\subsection{The DEEP1 Extended Sample}

Our sample consists of 101 galaxies, 93 of which are drawn from the DEEP1 
survey (see Vogt et al. 2004; Simard et al. 2002; Weiner et al. 2004), 
and 8 of which were obtained independently.  
Each of these systems has a resolved rotation curve which was
obtained with the Keck telescope using LRIS (Oke et al. 1995) 
in the redshift range $z \sim 0.2$ to $z \sim 1.2$. The additional data 
presented in this paper, which enables stellar masses to be compared with
virial masses, consists of near-infrared imaging described in
$\S$2.2.

Full details of the selection criteria for the galaxies are discussed
in DEEP1 papers (Vogt et al. 1996, 1997, 2005) to which the interested
reader is referred. Moreover, the necessary assumptions
implicit in the derivation are detailed in those papers, to which we
refer the reader.   Briefly, galaxies were selected morphologically
as elongated
disks in Hubble Space Telescope (HST) F814W (I$_{814}$) images with
I$_{814} < 23$. The
inferred inclination was chosen to be greater than 30\deg to
facilitate a measurement of the rotational velocity.
The optical images used for both photometric and morphological
analyses come from HST Wide Field Planetary
Camera-2 (WFPC2) observations of the Groth Strip (Groth et al.
1994; Vogt et al. 2004), the Hubble Deep Field (Williams et al. 1996), and 
CFRS fields (Brinchmann et al.
1998). Using I$_{814}$ images, structural parameters were
determined for each galaxy using the GIM2D and GALFIT packages (Simard et al.
2002; Peng et al. 2002). We fit a two-component model to the surface brightness
distribution, assuming a de Vaucouleurs law for the bulge, and an
exponential for the disk component.  Based on these fits, the disk scale
length, R$_{\rm d}$, and bulge-to-disk ratio ($B/D$) were determined.  The
uncertainties in the R$_{\rm d}$ values determined through this method are 
possibly underestimated, as has been explored through careful 2-D fitting
(e.g. Jong et al 1996). To be inclusive of possible
effects, we will incorporate an additional 30\% uncertainty
for our overall error on the measured values of $R_{\rm d}$.

Details of the observations, reductions and extraction of maximal
rotational velocities $V_{\rm max}$ from the LRIS spectroscopy are
presented in Vogt et al. (1996, 1997, 2005). Briefly, each disk
galaxy was observed along its major axis, as determined from the
HST images (Simard et al. 2002). The maximum velocity, $V_{\rm max}$, is 
determined by fitting a fixed form for the rotation curve scaled according to
the I$_{814}$ disk scale length $R_{\rm d}$. The assumed rotation
curve has a linear form which rises to a maximum at 1.5$\times$ R$_{\rm
d}$, and remains flat at larger radii. 
Our assumption that 1.5$\times$ R$_{\rm d}$ is the radius 
where the rotation curves for disks reach their maximum is reasonable, based
on a similar behavior for local disk systems of similar
magnitudes (Persic \& Salucci 1991; Sofue 
\& Rubin 2001).  Our rotation curves are also visible out to several scale
lengths, or roughly 2-5\arcsec\, (e.g., Vogt et al. 1997), adequate for
measuring V$_{\rm max}$.   This
model form is then convolved with a seeing profile that simulates the
conditions under which the observations were taken, and V$_{\rm
max}$ is determined by iterative fitting. Effects from the width
of the slit, slit misalignment with the major axis, and varying
inclination were taken into account when performing these fits, and
calculating the resulting errors.  Typically, $V_{\rm max}$ is
determined to a precision of 10-20\% due to these various sources
(e.g., Vogt et al. 1997).

\begin{figure*}
\begin{center}
\vspace{0cm}
\hspace{-1cm}
\rotatebox{0}{
\includegraphics[width=1\linewidth]{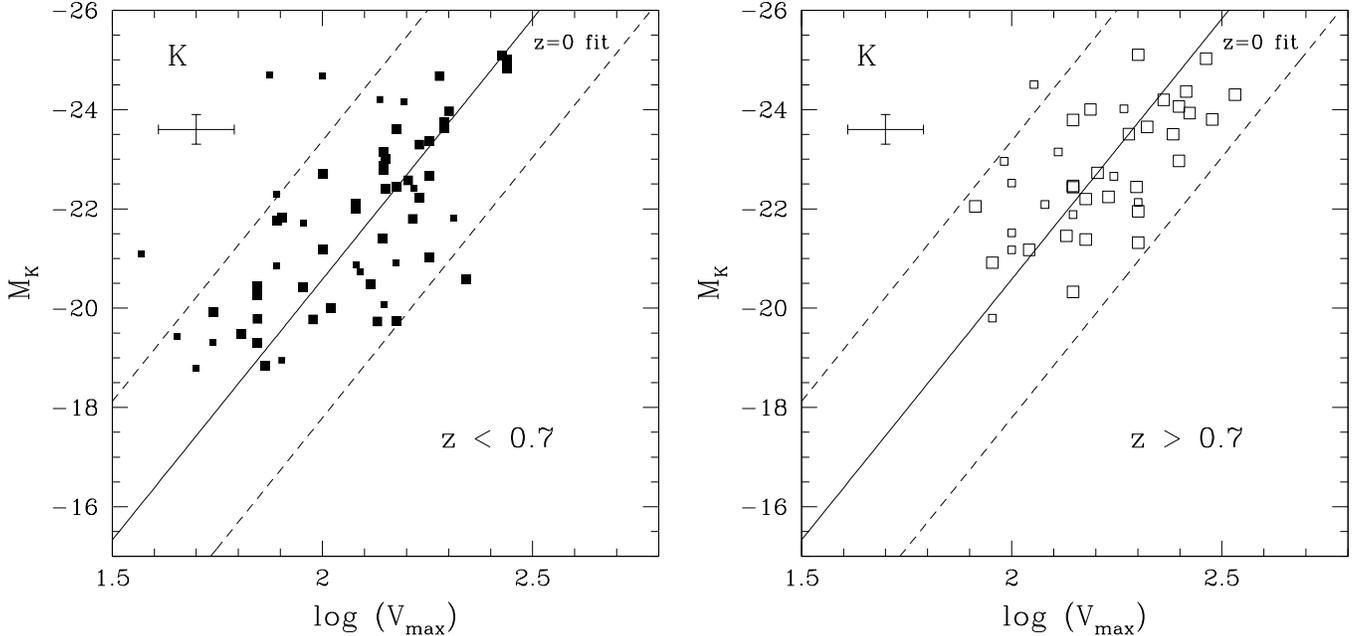}}
\end{center}
\vspace{-0.5cm}
\figcaption{The rest-frame K-band Tully-Fisher
relation for our sample of disks.  The panels are divided
into different redshift bins, higher and lower than $z = 0.7$.
The solid and dashed lines is the $z \sim 0$ Tully-Fisher
relation found by Verheijen (2001) and its $\pm$ 3$\sigma$
scatter. The average error is also plotted in each panel.  The
large points have errors lower than this average, while smaller
points have errors larger than the average.}
\end{figure*}

\subsection{Near-Infrared Imaging}

The new data we present in this paper consist of deep near-infrared
observations of the DEEP1 extended sample.
Precision near-infrared photometry is the critical ingredient for
determining stellar masses for our sample (Brinchmann \& Ellis 2000). 
Photometry was acquired
with a $K_s$ filter with three different instruments: the Keck 
Near Infrared Camera (NIRC, Matthews \& Soifer 1994), the UKIRT
Fast-Track Imager (UFTI, Roche et al. 2003), and the Cooled
Infrared Spectrograph and Camera for OHS (CISCO, Motohara et al. 2002) on the
Subaru 8.4 meter telescope (see Bundy et al. 2004 for a more detailed
description of these data).  NIRC has a field of view of
38\arcsec\ and a pixel scale of 0.15\arcsec~pixel$^{-1}$. The equivalent
numbers for UFTI are: 96\arcsec\ field of view, with a pixel scale of 0.091 
\arcsec pixel$^{-1}$.  The CISCO camera has a field of view of 108\arcsec\
with a pixel scale of 0.105\arcsec~pixel$^{-1}$.  The typical
depths for these images is $K_{s}$ = 20.5-21 (Vega) with a typical seeing of
$\sim$ 0.8\arcsec.

In each case, the infrared data were taken with a dither pattern
whose step size exceeded the typical size of the galaxies of
interest. The data were reduced by creating sky and flat-field
images from sets of disregistered neighboring science frames.
Standard stars were observed for calibration purposes during the
observations. Some images taken in good seeing but through thin
cloud were subsequently calibrated in photometric conditions via
shallower exposures taken with the Wide Field Infrared Camera
(WIRC, Wilson et al. 2003) on the Hale 5 meter telescope.

\subsection{Rest-Frame Quantities}

We measure our photometry in the $K_{s}$-band, and HST I$_{814}$ and
V$_{606}$ bands, within a scale-length factor, either 1.5R$_{\rm d}$ or
3R$_{\rm d}$, both of which are generally large enough to avoid seeing 
effects from the $K_s$ imaging.
We then extrapolate the total magnitudes, within each band, out to infinite
radius by using the
fitted parameters for an exponential disk derived in the HST I$_{814}$
band.  To compare observables over a range in redshift, it is
necessary to reduce all measures to a standard rest-frame.
Galactic extinction corrections were applied using the formalism
of Schlegel, Finkbeiner \& Davis (1998) and internal extinction
was accounted for according to measured inclinations and
luminosities using the precepts of Tully et al. (1998). It is
debatable whether internal extinction corrections
derived for nearby spirals are applicable to higher redshift
disks. Our sample is mostly composed of systems with M$_{\rm B} >
-22$ where extinction could be a cause for concern.  However, direct
extinction measurements in moderate redshift disks, determined
through overlapping pairs, find a modest overall extinction
(White, Keel \& Conselice 2000).

To derive absolute magnitudes and rest-frame colors, we estimated 
non-evolutionary $k$-corrections. The K-band $k$-corrections were computed 
from spectral energy distributions determined alongside the stellar mass
fits.   We experimented with both the original algorithm developed
by Brinchmann \& Ellis (2000) and also an independent one
developed by Bundy, Ellis \& Conselice (2005, see $\S$3.2).
Both methods agree very well. The approach is similar
to that used by Vogt et al. (1996, 1997) where $k$-corrections
were calculated from model SEDs from Gronwall \& Koo (1995) based
on various star formation histories in the Bruzual \& Charlot
(1993) code.  We also note that our derived k-correction values are very 
similar to non-evolutionary $k$-corrections derived by
Poggianti (1997) using SEDs from Poggianti \& Barbaro (1996).

\section{Mass Estimators}

\subsection{Virial and Halo Masses}

The virial mass of a galaxy - that is, the combined
total of the dark, stellar and gaseous components
- is perhaps its most fundamental property.
However, obtaining an accurate estimate from observed
quantities is necessarily difficult since there is
no {\it a priori} agreed model for the relative distributions
of the various components. Our approach to this
challenge will be to use both analytical techniques and
semi-analytical simulations to  investigate the relationship
between the total halo mass and  our dynamical and structural
observables. Although necessarily approximate and
debatable in terms of the assumptions made, we will attempt,
where possible, to investigate the uncertainties involved
by contrasting the two approaches in the context of
our data.

For a simple virialized system, such as a circularly rotating disk, we can
place constraints on the total mass within a given radius $R$ independent
of model assumptions. If $R > {\rm 1.5 \times R_{d}}$, where we estimate
the maximum rotational velocity V$_{\rm max}$ is reached, the mass within
$R$ is given by,

\begin{equation}
{\rm M_{vir}(<R)} = {\rm V_{max}^{2}}R/{\rm G},
\end{equation}

\noindent where $R > {\rm 1.5 \times R_{d}}$.  This assumes that the
rotation curve reaches the maximum velocity by 1.5 $\times$ R$_{\rm d}$,
(Persic \& Salucci 1991), otherwise V$_{\rm max}$ should be replaced
by V$(R)$.

There are a few possible approaches for determining total halo masses,
some of which require the use of simulations to convert observed dynamical
qualities, usually V$_{\rm max}$, into halo masses. We take a basic
approach using eq. (1) to obtain the total mass with a given radius, and
take a suitably large total radius of 100 kpc to measure the total
halo mass.  This is often the extent of disk HI rotation curves,
and similar to the sizes of dark matter halos (e.g., Sofue \& Rubin 2001).
Another approach now being used (e.g., Bohm et al. 2004) 
has been proposed by van den Bosch
(2002). In this case, it is argued from analytic simulations of disk
galaxy formation that the quantity $M_{\rm vdB} = 10.9 \times {\rm M_{vir}
(R_{d})}$ gives, on average, the best empirical representation of the
virial mass for simulated disks. The zero-point of the relationship
between R$_{\rm d}$V$_{\rm max}^2$/G and virial mass is claimed to be
independent of feedback and independent of the mass of the halo (van den
Bosch 2002).

Semi-analytic models based on $\Lambda$CDM (e.g., Cole et al. 2000; Benson
et al. 2002; Baugh et al. 2005) suggest, however, that the ratio between
V$_{\rm max}^{2}$R$_{\rm d}$/G and halo mass is not as simple as the above
formalism implies. The latest {\it Galform} models from Baugh et al.
(2005) and Lacey (priv. communication) show that equation [1] and
van den Bosch (2002)
under-predicts the dark halo mass. These models show that the relationship
between the virial mass at R$_{\rm d}$ and the total halo mass changes as
a function of mass in the sense that the ratio is higher for lower mass
halos. Physically this can be understood if high mass halos have a larger
fraction of their baryonic mass in a hot gaseous phase that is not traced
by the formed stellar mass.  Using the {\it Galform} models to convert our
observables into a total halo mass is however potentially inaccurate
as these models cannot reproduce the Tully-Fisher relation to 
better than 30\%.  However, there are reasons to believe that the semi-analytic
models put the correct amount of halo mass into their modeled galaxies
(e.g., Benson et al. 2000).   This however does
not necessarily imply that the V$_{\rm max}$ values in these models are
able to accurately match the halo masses.  Independent determinations of
total halo masses are necessary to perform this test.  However, the
masses from this approach match within 0.5 dex the masses from using
eq. (1) with a suitably large total radius.

With these caveats in mind, we have used the {\em Galform} model results to fit
the relationship between the virial mass at R$_{\rm d}$ (eq. 1), and the
total mass of the halo (M$_{\rm halo}$), a ratio which we call $\Re = {\rm
M_{vir}(R_{d})} / {\rm M_{halo}}$. We fit $\Re$ as a linear function of
${\rm M_{vir}(R_{d})}$, such that
$\Re = \alpha \times {\rm log (M_{vir}(R_{d}))} + \beta$.  Using
the {\em Galform} results we fit $\alpha$ and $\beta$ at redshifts $z = 0, 0.4,
0.8, 1.2$. We find that the functional form of $\Re$ does not change
significantly with redshift, with typical values $\alpha = -0.1$ and
$\beta = 1.3$.  The value of the halo mass M$_{\rm halo}$ is then given
by:

\begin{equation}
{\rm M_{halo}} = {\rm M_{vir} (R_{d})}/\Re.
\end{equation}

\noindent Observational and model uncertainties contribute to errors on 
these virial
and halo mass estimates in two ways. Measured scale lengths given by the
GIM2D and GALFIT fitting procedure (\S 2.1), give an average error of 0.12
kpc, although we add an extra error to this to account for
systematics seen when performing one component models to disks/bulge systems
(de Jong 1996). 
At the same time, systematic difficulties in the rotation
curve analysis (\S 2.1) can arise. Vogt et al. (2005) discuss these issues
in some detail and conclude the average error is $\sim 27$ \kms\,
(typically $\simeq$10-20\%). Following the discussion of rotation
curve fitting in $\S$2.1, it is possible we do not 
measure the true V$_{\rm max}$ required for insertion in equations
(1 \& 2). Even a modest underestimate of V$_{\rm max}$ would lead to
a significant error in $M_{\rm vir}$. In combination, these
measurement uncertainties imply virial and halo masses precise to no
better than 30\%.  The semi-analytic method for computing total halo
masses is likely limited, and thus we also investigate the total halo masses 
found through the van den Bosch (2002) formalism, and through the use of
equation (1) out to 100 kpc.  We find that the total halo mass through 
eq. (1), eq. (2) and
van den Bosch (2002) are all fairly similar.  We account for these
differences in our use of total halo masses in \S 4.3.

\begin{figure*}
\begin{center}
\vspace{0cm}
\hspace{1cm}
\rotatebox{0}{
\includegraphics[width=1\linewidth]{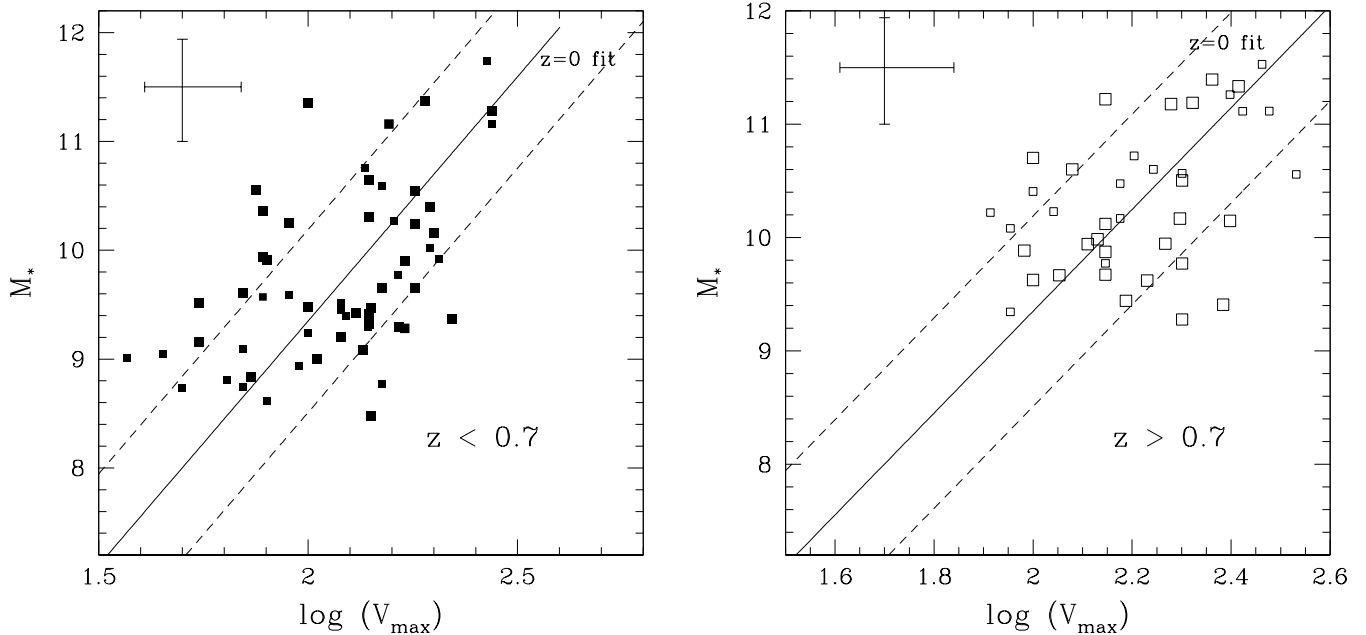}}
\end{center}
\vspace{0cm}
\figcaption{The stellar mass Tully-Fisher
relation plotted as a relation between M$_{*}$ and V$_{\rm
max}$.  The solid and dashed line is the $z = 0$ relationship
found by Bell \& de Jong (2001) for nearly disks and its $\pm
3\sigma$ scatter. The error bar is the average with large points
having errors lower than this average, and smaller points having
errors larger than the average.}
\end{figure*}
\subsection{Stellar Masses}

Our procedure for deriving stellar masses follows the multi-color
method introduced by Brinchmann \& Ellis (2000). Combining HST
V$_{606}$, I$_{814}$, and near-infrared $K_s$ photometry for a galaxy of 
known redshift, we fit a range of template SEDs synthesized using
software by Bruzual \& Charlot (2003).   Our photometry is
done in a matched aperture  large enough to avoid seeing problems
associated with the $K_s$-band.  
These fitted SEDs constrain the
$K_s$-band mass/light ratio.  Our code for computing stellar masses 
only uses models with ages less than the age of the
universe at the redshift observed. A $\chi^2$ analysis normalized by
the near infrared $K_s$-band luminosity yields the stellar mass.
The template SEDs were constructed sampling a range of
exponentially-declining star formation rates, metallicities and
ages with a Salpeter IMF.  Using other stellar mass functions, such
as Chabrier, Kennicutt or Kroupa would
result in computed stellar masses that are smaller by $<$ 0.3 dex.
We further assumed a simple exponentially 
declining star formation history with  $\tau$ values 
ranging from 0.001 - 15.0 Gyr, and metallicities  from Z=0.005-5 in
solar units. Typical uncertainties in this method are a factor 
of three (Brinchmann \& Ellis 2000; Papovich, Dickinson \& Ferguson 2001;
 Drory et al. 2004; Bundy, Ellis \& Conselice 2005).

The SED is constrained by the observed colors as measured in an
aperture of radius 1.5 $\times$ R$_{\rm d}$ in each band which is
optimal in terms of signal-to-noise (S/N).  We assume that there
are no color gradients and use the colors measured within this
aperture as the global color.  The total $K_s$-band light is measured
by extrapolating the 1.5 $\times$ R$_{\rm d}$ flux to infinity, assuming
the same exponential fit as measured in the HST I$_{814}$ image.

Errors on the stellar masses arising from photometric uncertainties
can be determined in a Monte Carlo fashion.  Simulated exponential disks 
of a known magnitude were inserted into the reduced HST and ground
based images and photometrically recovered using the tools that
were applied to the sample. The simulated disks were arranged to
randomly sample the selected ranges of disk scale-length $R_{\rm d}$ and
inclination. The derived photometric errors were then input into
the stellar mass calculations from which a $1 \sigma$ range was
calculated.  The average error in M$_{*}$ is 0.47 dex.

The calculation of stellar masses through this technique is limited to
a degree by systematics which are difficult to constrain with the
current data set, with $K_s$-band 
magnitude errors only a small source of uncertainty. 
The models of the spectral evolution of galaxies
depend both on the observed stellar libraries and underlying theory,
and some fundamental uncertainties remain (c.f.\ Bruzual \& Charlot
2003).  We expect uncertainties of about 5-10\%,
given the range of possible models (Charlot, Worthey \& Bressan
1996). This is not important for the conclusions in this paper but
might be an issue with larger and more accurate datasets in the
future.

We also cannot constrain the amount of recent star formation produced
in bursts (e.g., Kauffmann et al.  2003; Bell \& de Jong 2001),
resulting in a slight systematic overestimate of the stellar
mass. The latter effect only becomes important at large burst
fractions and should be $<10$\% for our big spirals (c.f.\ Drory et al
2003). The final systematic uncertainty is the dust correction
adopted. We use the Calzetti (1997) extinction law in
our stellar mass calculations, although other extinction laws produce
very small differences (at most 10-20\%) in the resulting stellar mass
(Papovich et al. 2001).  Taken together we are likely to have systematic 
uncertainties in our
stellar mass estimates amounting to $\sim 0.15$ dex which is lower
than our random uncertainty, thus they will not influence our results.

\section{Results}

\subsection{K-band Tully-Fisher Relation}

The Tully-Fisher (TF) relation has been the traditional method for
investigating how dark halos and the stellar components of disk
galaxies relate.  The optical relation has been studied in disks
galaxies out to redshifts $z \sim 1$ by e.g., Vogt et al. (1996,
1997, 2005), Ziegler et al. (2002) and Bohm et al. (2004).  These
investigations have found between 0.4 and $\sim 1$ magnitudes of
rest-frame B-band luminosity evolution in disks between $z \sim 0$
and $z > 0.5$.  This luminosity evolution is derived by assuming
that the slope of the TF relation at high redshift is the same as
it is at $z \sim 0$.  The question of differential evolution in
the relation (Ziegler et al. 2002; Bohm et al. 2004), for
example that the faintest galaxies evolve more rapidly, remains an
important unknown.

Although the purpose of this paper is to move beyond the TF
relation, we begin by plotting the K-band TF relation for our sample 
(Figure~1).  One might expect evolution of the TF relation
in the K-band to display a clearer signal than the B-band since the effects of
dust are mitigated and passive evolution should be more uniform in
its effect across the sample.  When we assume that the slope of the
K-band Tully-Fisher is the same as the local value from Verheijn (2001)
we find no significant evolution, as is also found in the B-band
(Vogt et al. 2005).  We perform these fits using both a downhill
simplex amoeba and Levenberg-Marquardt $\chi^{2}$ minimization, both
of which give the same results.

We find a fading of 0.04$\pm$0.24
magnitudes for systems at $z > 0.7$ compared with the $z \sim 0$
relationship and a brightening of 0.37$\pm$0.23 magnitudes for systems
at $0.2 < z < 0.7$, consistent with no evolution.   The scatter does not
evolve significantly (1.13 magnitudes for the $z<$0.7
sample, and 0.72 magnitudes for systems at $z > 0.7$).  In all cases, the 
observed
slope and scatter change only slightly when we ignore internal
extinction corrections.  

Although our results are broadly consistent with earlier, smaller
samples, the interpretation of any evolutionary signal is
complicated in two ways. First, only a limited range of
luminosity and rotational velocity can be sampled at high
redshift, leading to great uncertainties given the intrinsic
scatter. Second, as luminosity and rotational velocity are
indirect measures of the assembly state of the galaxy, both may be
evolving in complex ways, that mask actual evolutionary changes.

\subsection{The Stellar Mass Tully-Fisher Relation}

The first step beyond the TF relation is to compare the stellar
mass to the measured maximum velocity - a relation we will call
the {\em stellar mass - Tully Fisher relation}.  The classical
B-band TF relation scales such that $L\, \propto\, V^{3.5}$.  This
coupling becomes even steeper for the local stellar mass
Tully-Fisher relation in nearby disks, $M \propto V^{4.5}$ (Bell
\& de Jong 2001).

Ideally, we seek to measure the all-inclusive baryonic TF
relation, but measuring the gas content of high redshift disks is
not yet feasible.  We can estimate how much cold gas we are
missing in our stellar mass inventory by investigating the gas
mass fractions for nearby disk galaxies. Through examinations of
the luminosities and HI masses for nearby disks, McGaugh \& de
Blok (1997) conclude that galaxies which are massive, bright,
red, or have a high-surface brightness, have very little gas in
comparison to bluer, fainter, lower surface brightness systems.  
Systems which are brighter than M$_{\rm B} = -21$ have gas mass
fractions that are typically 0.1 or lower.
Since our selection finds the most luminous, high surface
brightness systems, which are also red, they are the least likely
sub-class of disks to have a high gas content.

\begin{inlinefigure}
\begin{center}
\vspace{2cm}
\hspace{-0.5cm}
\rotatebox{0}{
\resizebox{\textwidth}{!}{\includegraphics[bb = 25 25 625 625]{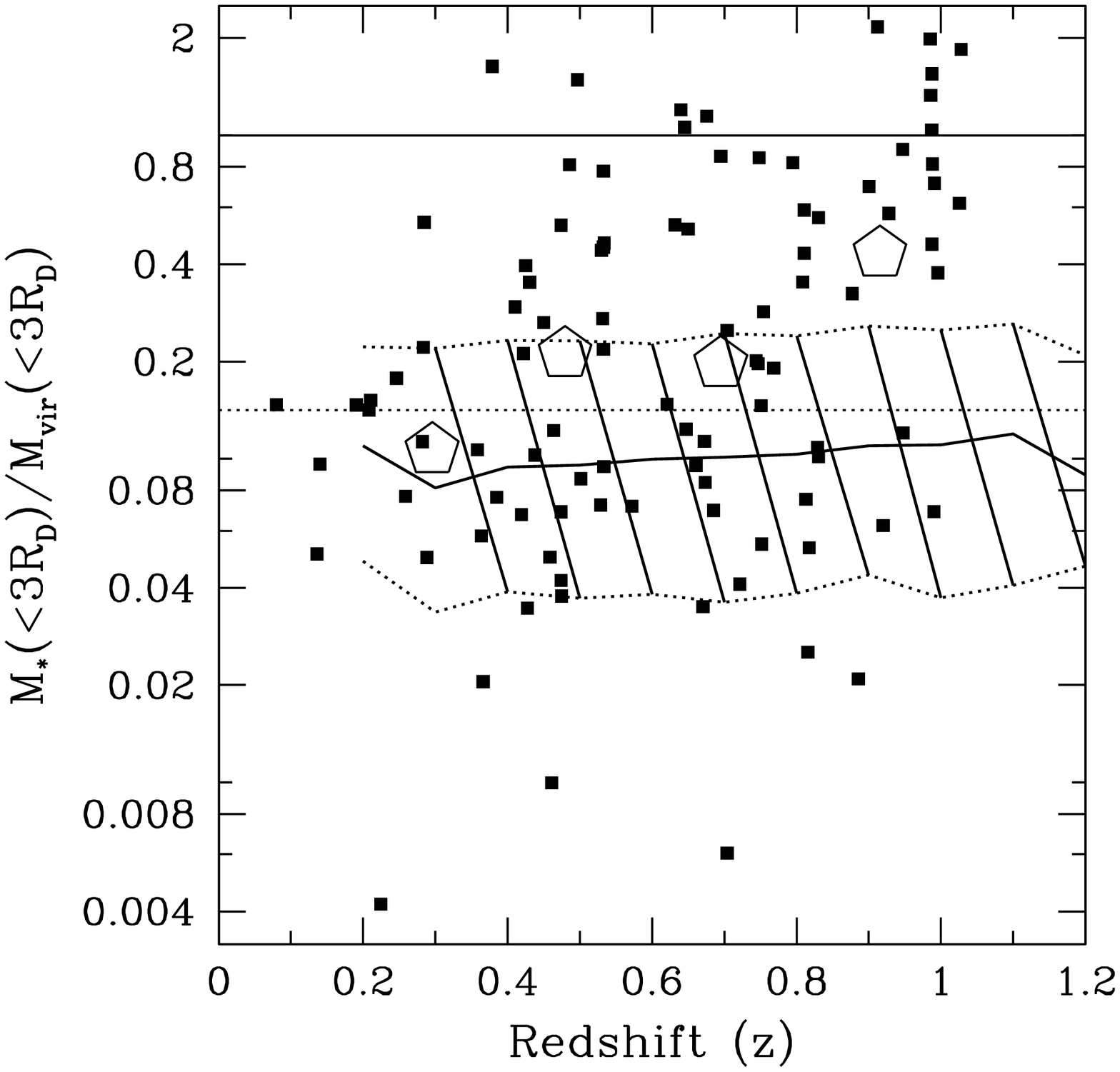}}}
\end{center}
\vspace{-2cm}
\figcaption{The relationship between the stellar mass and virial
mass within 3R$_{\rm D}$ plotted as a function of redshift ($z$)
for all galaxies in our sample.  The pentagons show the average
value of this ratio as a function of redshift. The solid horizontal line shows
the location of M$_{*}(<{\rm R_{d}})$ / M$_{\rm vir}(<{\rm R_{d}})$ = 1
while the horizontal dotted line is the universal baryonic mass limit.  
The solid line surrounded by the hatched region shows the predictions of 
a hierarchical $\Lambda$CDM based galaxy formation model (Baugh et al. 2004). }
\end{inlinefigure}

The stellar mass TF relation is shown in Figure~2 where, as before,
we divide the sample into two redshift bins, split at $z = 0.7$.
Each panel contains a solid line giving the $z \sim 0$ best fit
and a dashed line illustrating the $\pm$ 3 $\sigma$ uncertainty in
this fit (Bell \& de Jong 2001).  As was the case for the
conventional TF relations, no significant evolution in the
zero-point is observed. The Bell \& de Jong (2001) $z = 0$ stellar
mass Tully-Fisher relation can be written as M$_{*} = 0.52 + 4.49 \times
{\rm log (V_{max})}$.  By holding the slope of this relationship constant,
we find that the zero point is best fit by 0.45$\pm$0.12 at $z < 0.7$
and 0.41$\pm$0.13 at $z > 0.7$.  Neither of these are however
significantly different from the $z \sim 0$ relationship, and are very similar
to each other. This implies that, if growth continues, the stellar and dark 
components are
growing together. For example, if disk assembly since $z\simeq0.7$
proceeded only by the addition of stellar mass at a uniform
rate of 4 M$_{\odot}$ year$^{-1}$, the local zero point would be
discrepant at the 4$\sigma$ level.
This lack of evolution is important for understanding how disk
galaxy formation is occurring (see \S 4.4).

Moreover, the scatter in the stellar mass Tully-Fisher relation 
is similar to that
observed in the K-band Tully-Fisher after converting the K-band
magnitude scatter into a luminosity and assuming an average
stellar mass to light ratio.  The typical scatter (in log M$_{*}$ units)
in stellar mass for these is 0.65 for disks at $z < 0.7$ and 
0.48 for those at $z > 0.7$.

\begin{figure*}
\begin{center}
\vspace{0cm}
\hspace{-1cm}
\rotatebox{0}{
\includegraphics[width=1\linewidth]{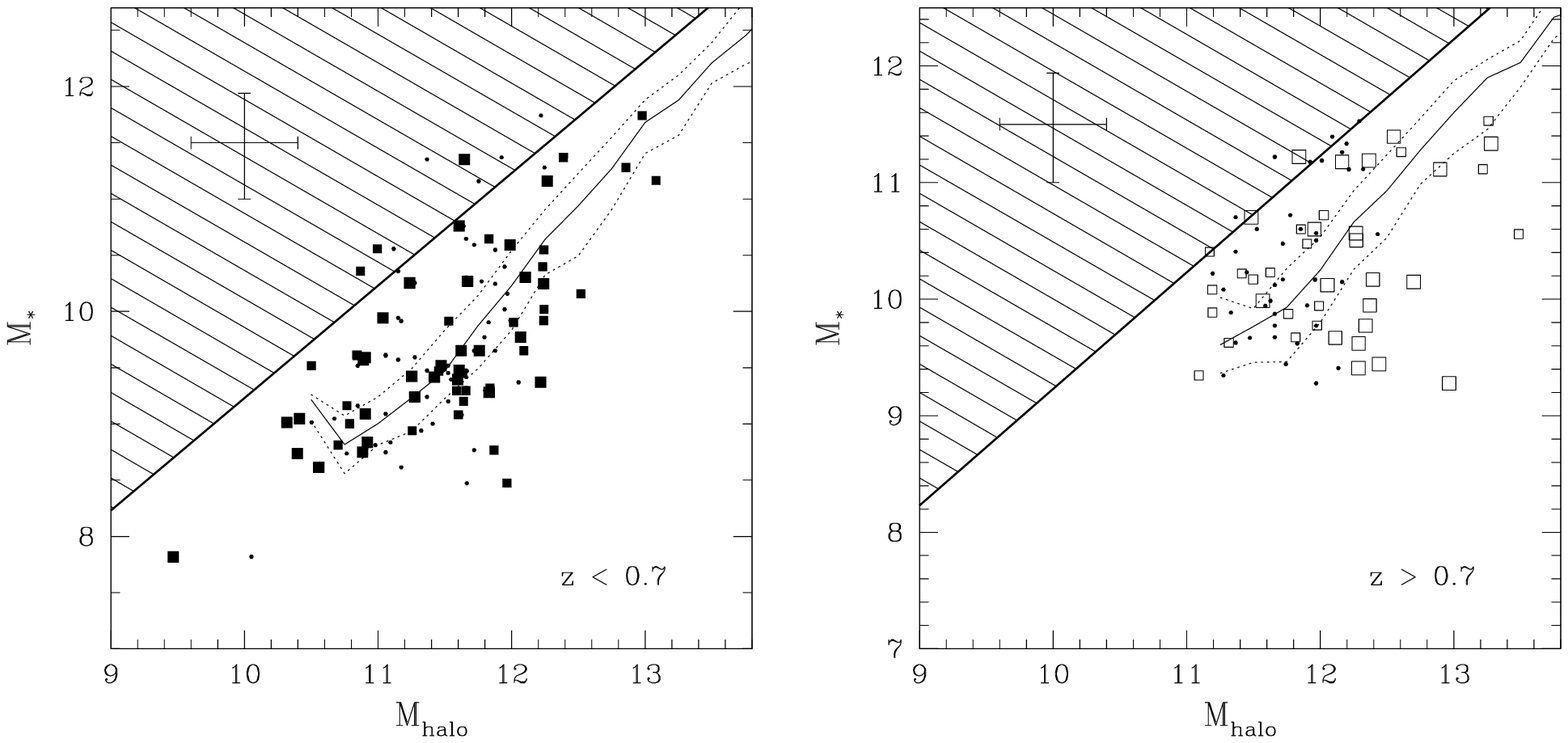}}
\end{center}
\vspace{-0.5cm}
\figcaption{The relationship between stellar mass and
halo masses in our two different redshift bins.  The large symbols
are for total masses derived from model relationships (eq. 2), 
and the small points are total halo masses derived using eq. (1) with 
$R$ = 100~kpc. The thin solid line is the relationship
between stellar and virial masses from the semi-analytic models of
Benson et al. (2002) at $z = 0.4$ for the $z < 0.7$ sample and at $z
= 0.8$ for the $z > 0.7$ sample. The short dashed lines display
the 80\% range of where galaxies in these simulation are found. The thick solid
line is the baryonic fraction limit and the shaded region is the
area where the stellar mass fraction is greater than the universal
baryonic mass fraction. }
\end{figure*}

\subsection{A Comparison of Stellar and Halo Masses}

The final step in our analysis will be an attempt to  
convert
our measured quantities into a comparison of the stellar
and halo masses as discussed in $\S$3. Recognizing the
considerable uncertainties involved, Figure 3 shows as a function
of redshift the ratio, $f_{*} =$ M$_{*}$/M$_{\rm vir}$,
between the stellar masses and virial masses (eq. 1) within 
3R$_{\rm d}$ for our sample.     As expected, the stellar masses 
are nearly always less than the
independently-derived virial masses, indicating
that our methods for computing these values are not dominated by
large systematic errors.  It can also be seen from
this figure that there is wide range of M$_{*}$/M$_{\rm vir}$ values
at every redshift from $z \sim 0.2 - 1.2$. The open pentagons show
the median values of $f_{*}$ as a function of redshift.   The dotted horizontal
line shows the global baryonic mass fraction $\Omega_{\rm
b}/\Omega_{\rm m}$ = 0.171 derived from WMAP results (Spergel et
al. 2003).  Within 3R$_{\rm d}$ it appears that many disks have
M$_{*}$/M$_{\rm vir}$ values higher than this limit.  It also appears
that within 3R$_{\rm d}$ the stellar mass of the disk dominates
the virial mass.   This indicates that within the visual parts of some of 
our sample, the disk component accounts
for roughly all the mass, which is consistent with a
maximal disk interpretation.   However, there are clear examples on
Figure~3 where either the stellar mass is overestimated and/or the
virial mass is underestimated. Both of these are possibilities, since
our stellar masses are potentially too high from using a Salpeter IMF, and
we might be underestimating the value of V$_{\rm max}$ due to seeing, or
a lack of depth in the LRIS observations.  There is also a slight bias
such that at higher redshifts nearly all galaxies sampled have high
V$_{\rm max}$ values, which are likely maximal, while at lower redshifts
we are sampling systems with lower V$_{\rm max}$ values, that have not yet
fully formed their stellar masses.  This is likely part of 
the reason that the stellar
mass to virial mass ratio increases slight at higher redshifts.

However, we are also interested in constraining the relationship
between the total halo mass, M$_{\rm halo}$, and the total
stellar mass M$_{*}$. As we discuss in \S 3, it is very difficult
to accurately obtain halo masses.  We therefore show in Figure~4 total
halo masses derived through eq. (1) at 100 kpc and through the
semi-analytical approach (eq. 2).   Although we cannot
accurately determine total halo masses  for individual systems,
our main goal is to compare how the ratio of stellar to halo mass changes
with redshift.  Figure~4 shows this
relationship divided into the same redshift bins as in Figures 1 \& 2.
To first order, the halo masses and stellar masses of disk galaxies should
correlate if star formation is regulated in the same manner in halos
of different masses (Steinmetz \& Navarro 1999).
Although there is no reason to expect any particular functional
form between these two quantities, there is a reasonably well-fit
linear relationship between them. We find that the zero point and slope of 
this relation does not change
significantly between low and high redshift.  There is also no
obvious change in the scatter from high to low
redshift (0.26 cf. 0.32).   The most significant
outcome of Figures 3 and 4 however is the remarkable similarity in trends
found at high and low redshift suggesting some disks had completed
the bulk of their stellar assembly by $z\simeq$1 or more likely that
the stellar and dark masses of galaxies grow together continuously.

Figure~5 shows the distribution of $f_{*}$ = M$_{*}$/M$_{\rm halo}$ 
for systems more massive than, and less massive than, the average halo mass,
M$_{\rm halo} = 10^{11.8}$ \solm. The solid line is the universal
baryonic mass ratio. As mentioned earlier, most disk stellar mass fractions 
are lower than the cosmic
ratio, which seems appropriate given our stellar mass inventory is
not intended to account for all associated baryons. 
Although some dispersion is
expected, conceivably some fractions are overestimated or underestimated, for
example by making incorrect assumptions about the IMF, or
underestimating V$_{\rm max}$ (and hence M$_{\rm halo}$) by
insufficient sampling of the rotation curve.

Figure~5 also shows a tentative population of disk galaxies with
remarkably low stellar fractions, the most extreme cases occurring
in objects with halo masses M$_{\rm vir} > 10^{11.8}$ \solm.
These galaxies deviate from the $z = 0$ stellar mass TF
relation by more than 4 $\sigma$. Investigation of the
individual systems that lie in this category shows them to be
undergoing vigorous star formation as inferred by bluer than
average $(U-B)$ colors. A weak correlation was found
between $f_{*}$ and $(U-B)$.   Because these systems are blue, they
may have kinematic asymmetries in their rotation curves which may raise
their V$_{\rm max}$ values. However, this is unlikely the effect producing
this slight correlation as most of the low $f_{*}$ systems also
have low V$_{\rm max}$ values. We investigated several other possible
correlations involving $f_{*}$ (for example with the bulge/disk
ratio) but no significant trends were found.

\subsection{Comparison with Models}

To investigate the implications of our results for the assembly
history of stellar mass in disk galaxies since $z \sim 1$, we
return to the behavior of the stellar fraction $f_{*}$ =
M$_{*}$($<$3R$_{\rm d}$)/M$_{\rm vir}$($<$3R$_{\rm d}$) vs. redshift
(Figure~3), and the relationship between total stellar mass (M$_{*}$) and
halo mass (M$_{\rm halo}$) at $z > 0.7$ and $z < 0.7$ (Figure 4). Trends in 
these relations may shed light on the physical processes regulating
star formation in disk galaxies. This is only the case however if the
value of R$_{\rm d}$ for disks does not grow with time, which appears
to be the case for the largest disks, based on our own limited data, and in 
statistical studies of $z > 0.8$ disks (Ravindranath et al. 2004; Conselice 
et al. 2004; cf. Ferguson et al. 2004 for higher redshifts).

First, Figure~3 shows that the median stellar vs. virial mass value for
our sample does not change significantly with redshift. We can compare
two simple extreme models to this result to determine the likely method
by which disk galaxies are forming at $z < 1$.
In a `monolithic collapse' model, 
where the dark halo and baryons for a galaxy are in place at high redshift,
and there is no change in these mass components with time (e.g., Eggen,
Lynden-Bell \& Sandage 1962), the average value of  M$_{*}$/M$_{\rm vir}$ 
should increase with time.  This assumes that effective radii distribution
for disks does not change with time, which is observed (Ravindranath et al.
2004).  There is evolution
in the monolithic model only in the sense that baryons are gradually 
converted to stars over time. This is the opposite to the observed trend in
which the ratio of  M$_{*}$/M$_{\rm vir}$ is roughly constant.

On the other hand, in a hierarchical picture, the value of 
M$_{*}$/M$_{\rm vir}$ remains relatively constant with time.  Shown
in Figure~3 are model predictions from {\em Galform} 
which show that the value of M$_{*}$/M$_{\rm vir}$ remains relatively
constant with redshift.  While we do not have a complete sample, 
we do find that the average and median values of 
M$_{*}$/M$_{\rm vir}$ for our sample remain constant with redshift within
the errors, which is consistent with the hierarchical idea. The higher
average M$_{*}$/M$_{\rm vir}$ value at $z > 0.8$ is likely produced
by our selection of the brightest disks at these redshifts (Vogt et al. 2005). 
The {\em Galform} model however does not predict the full range of mass 
fractions 
that we find. At all redshifts there are systems dominated within R$_{\rm d}$ 
by their stellar masses, while some systems have low M$_{*}$/M$_{\rm vir}$
ratios.  This is either produced by an observational bias (\S 4.3), or
the simulations predict too much dark matter in the centers of disks.

\begin{inlinefigure}
\begin{center}
\vspace{2cm}
\hspace{-0.5cm}
\rotatebox{0}{
\resizebox{\textwidth}{!}{\includegraphics[bb = 25 25 625 625]{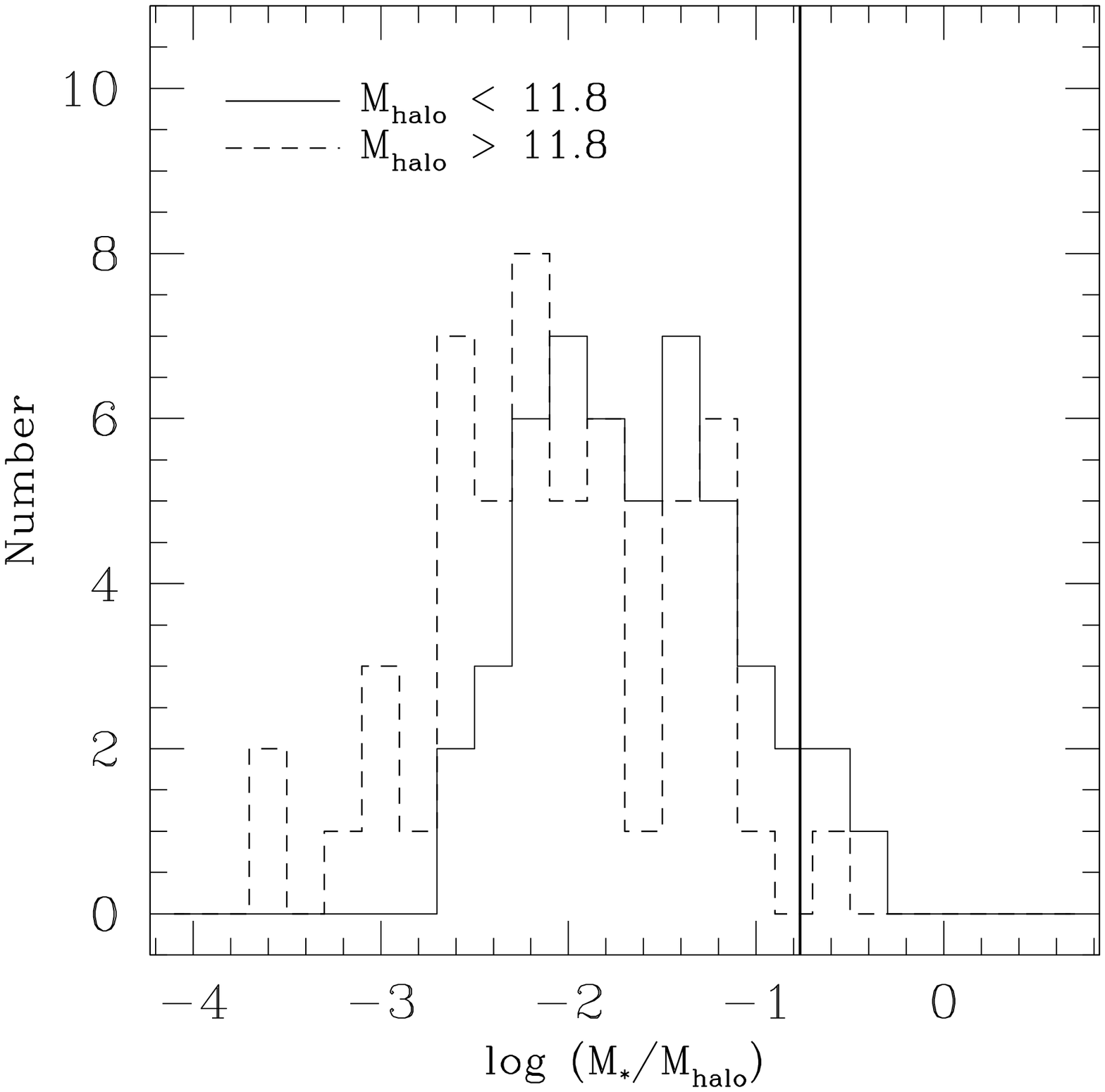}}}
\end{center}
\vspace{-2cm}
\figcaption{Histogram of M$_{*}$/M$_{\rm halo}$ values
divided into disks of different virial masses.  The solid line is
the global baryonic mass fraction.}
\end{inlinefigure}

Figure~4 shows the relationship between total stellar and total halo
masses (see \S 3.1) for our sample, divided into our two redshift ranges.  We 
also show on Figure~4 the {\em Galform} semi-analytic model values for
this relationship for disk dominated galaxies with the
80\% completeness indicated.  As in Figure~3, no strong evolution in
M$_{*}$/M$_{\rm halo}$ ratios is predicted. As galaxies in the
semi-analytic models grow by accreting smaller systems, or intergalactic
gas coupled with dark matter, and then
converting the newly-obtained gas into stars quickly, the
relationship between M$_{*}$ and M$_{\rm halo}$ remains constant.  This
is generally what we find, as well as a good agreement with the model
predictions.  In fact, the agreement between the total stellar and halo
masses with the {\em Galform} models is slightly better than the comparison
between
the stellar and virial masses with 3$\times$R$_{\rm d}$, which further
suggests that the luminous components of disk galaxies are dominated by
the stellar mass (\S 4.3).

In summary, the fact that the ratio  M$_{*}$/M$_{\rm halo}$ remains relatively
constant in our sample with redshift, and that the stellar-mass TF relation
does not evolution, are indications that disk galaxies 
are forming through the accretion of both dark and baryonic mass.  
Disk galaxies are undergoing star formation at $z < 1$ at a rate
of a few solar masses per year. If disks
at $z \sim 1$ contained all the  baryonic or halo masses that they
have at $z \sim 0$, we would see an increase in 
M$_{*}$/M$_{\rm halo}$ with time.   Because we do not see this trend,
it appears that new stars form out of gas accreted from the intergalactic 
medium, which is coupled with dark matter at a constant ratio.  This baryonic 
to dark matter ratio is such that both 
M$_{*}$ and M$_{\rm halo}$ grow with time, creating a stellar mass
to halo mass ratio that is on average relatively constant.

\section{Conclusions}

We present the results of a dynamical and structure study
of 101 disk galaxies drawn mostly from the DEEP1 survey
with redshifts in the range $z\simeq$0.2-1.2.   New infrared
imaging observations are presented which enable us to
derive reasonably reliable stellar masses and thereby to
construct the stellar mass Tully-Fisher relation and its
redshift dependence. Using various formalisms drawn
from analytic and semi-analytical models, we attempt to
convert our dynamical data to make the first comparisons
of the relative fractions of stellar and total mass in our sample.
Notwithstanding the considerable uncertainties and
sample incompleteness, the results are encouraging
and suggest remarkably little evolution in the mix of
baryons and dark matter since $z\simeq$1.

Although our sample is not formally complete in luminosity or mass, 
we explore the degree to which there may have
been evolution in the relative distribution of stellar and virial
masses and find the following:

\begin{enumerate}

\item{} Massive disk galaxies exist out to $z \sim 1$ with halo
 masses as large as 10$^{13}$ \solm, roughly as large as
the most massive disks in the nearby universe.  These systems also
contain a large amount of stellar mass.  At least some disk
galaxies are nearly fully mature in their stellar content at $z \sim
1$.

\item{} We confirm earlier studies based on smaller samples and
find no significant evolution in the zero point or scatter of the
rest-frame $K$-band Tully-Fisher relation out to $z \sim 1.2$.

\item{} The stellar mass Tully-Fisher relation out to $z \sim 1.2$ is
likewise largely consistent with the relation found for nearby
disks. We find no significant evolution in our sample after
comparing systems at redshifts greater than and less than $z =
0.7$.

\item{} Although there are clearly great uncertainties
in estimating total halo masses from our dynamical
data, we find the distribution of the ratio of stellar and halo masses
remains relatively similar from $z \sim 0$ to $z \sim 1.2$. The
stellar fraction observed can be understood if the bulk of the
baryons associated with massive disk galaxies have already formed
their stars. A modest number of massive galaxies have very low
stellar fractions, consistent with continued star formation as
revealed by their blue $U-B$ colors.

\item{} These results are in relatively good agreement with $\Lambda$CDM 
analytical
and semi-analytical models (Benson et al. 2002; Baugh et al. 2005), 
suggesting that disk galaxy formation is hierarchical in nature.

\end{enumerate}

Our overarching conclusion from this study is that no significant
evolution in the stellar mass fraction can be detected in the
population of regular massive disks since $z\simeq$1. Although
biases and uncertain assumptions may affect detailed quantities at
the $0.3-0.5$ dex level, the absence of gross trends is consistent
with the conclusion that the bulk of these systems grow at $z < 1$
by the accretion of dark and baryonic material.  This conclusion is
however tempered by the fact that our selection is affected by
a bias that likely changes with redshift.  For example, because our sample
was selected by a $I < 23$ limit, we
are studying the brightest disks at the highest redshifts.

These results, while intriguing, will be examined in much greater
detail with the upcoming DEEP2 redshift survey (Davis et al. 2002),
which will combine near-infrared imaging from the Hale 5 meter
telescope with rotation curves obtained using the DEIMOS
spectrograph on the Keck II telescope.  There are several issues
that can be better addressed with deeper spectroscopy and a more
careful selection of high-z disks.  The total number of disks in
this much larger sample is expected to be on the order of
thousands and will solidify and expand on the preliminary results
revealed by this study.

We thank the DEEP team for generously enabling us to augment their optical
sample and catalogs with near-infrared data in order to make this comparison.
We also thank Cedric Lacey and Andrew Benson for access to their
semi-analytical
model simulation results, and Xavier Hernandez and Ken Freeman for
comments regarding this work.   CJC acknowledges support from a National
Science Foundation Astronomy and Astrophysics Postdoctoral
Fellowship. NPV is pleased to acknowledge support from NSF grants 
NSF-0349155 from the Career Awards program, NSF-0123690 via the Advance 
IT program at NMSU, and AST 95-29098 and 00-71198 administered
at UCSC, and NASA STScI grants GO-07883.01-96A, AR-05801.01, AR-06402.01,
and AR-07532.01.

\newpage

\end{document}